\documentclass[11pt]{article}

\usepackage{hyperref,amsfonts,amsmath,amssymb,bm,a4wide,cite,graphicx}

\numberwithin{equation}{section}

\begin{document}

\title{\textbf{Effective propagators of flavor neutrinos}}

\author{Maxim Dvornikov\thanks{maxim.dvornikov@gmail.com}
\\
\small{\ Pushkov Institute of Terrestrial Magnetism, Ionosphere} \\
\small{and Radiowave Propagation (IZMIRAN),} \\
\small{108840 Moscow, Troitsk, Russia}}

\date{}

\maketitle

\begin{abstract}
We analyze the possibility to construct propagators of flavor neutrinos,
which are particles with indefinite masses. This kind of propagators
was used previously while studying neutrino flavor oscillations in
frames of the quantum field theory (QFT) based approach. Starting
with the operator and the path integral formulations of the QFT, we
obtain that the naive derivation of propagators for flavor neutrinos
violates the basic principles of the QFT. Nevertheless, we can still construct
the effective propagator of flavor neutrinos considering the mass
mixing term as a perturbation. In this situation, the effective propagator
arises from the solution of a Dyson-like equation. We use the derived
effective propagator to rederive the quantum mechanical transition
probability for neutrino oscillations in vacuum in frames of the QFT.
The application of the obtained results to neutrino oscillations in
background matter is also considered.
\end{abstract}

\section{Introduction}

Neutrinos were experimentally confirmed, e.g., in Refs.~\cite{Fuk98,Ahm02},
to be massive particles with a nonzero mixing between different particle
types. It means that flavor eigenstates, which interact with charged
leptons in the standard model, are the superposition of mass eigenstates,
which the fundamental particles. The neutrino mixing results in neutrino
flavor oscillations, which are the periodic change of the flavor content
in a neutrino beam.

Flavor oscillations are described mainly in frames of the quantum
mechanics (QM) (see, e.g., Ref.~\cite{Bil15}). The QM approach provides
satisfactory description of neutrino oscillations in the majority
of practical cases. Nevertheless, to develop the comprehensive description
of flavor oscillations, the quantum field theory (QFT) has to be applied
to the problem in question. It should be noted that the idea that
the QFT should be used for neutrino oscillations has been realized for
quite some time ago (see, e.g., Refs.~\cite{Beu03,NauNau20}, where
numerous approaches have been reviewed).

We adopt the QFT approach for flavor oscillations where neutrinos
are virtual particles, which was developed in Refs.~\cite{Kob82,GiuKimLee93,GriSto96}
originally. Since neutrinos are virtual in this method, their contribution
to the matrix element is concentrated in a neutrino propagator. Thus,
the milestone of this approach is the derivation of a neutrino propagator.
Note that, besides flavor oscillations in vacuum, this kind of formalism
was used to study oscillations in external fields in Refs.~\cite{CarChu99,AkhWil13,EgoVol22,Dvo25,Dvo26a,Dvo26b}.

It should be noted that, sometimes, the features of fundamental mass
eigenstates are borrowed for flavor ones, which are the way to express
the Lagrangian of the interaction of fundamental mass eigenstates
with charged leptons in the standard model. However, because of the
smallness of neutrino masses, this kind of borrowing does not affect
the final result, which is the probability of neutrino oscillations.

We mentioned earlier that the QFT approach to flavor oscillations
involves neutrino propagators. In Refs.~\cite{CarChu99,AkhWil13,KriSim23,KovSim24},
while using the QFT approach for flavor oscillations, the propagators
of flavor neutrinos were defined in the similar manner as for fundamental
mass eigenstates. Despite the final transition probability of oscillations,
derived in Refs.~\cite{CarChu99,AkhWil13,KovSim24}, turned out to
be correct, one should find out whether the theoretical input in these
works is justified. The aim of the present work is to examine the
validity of the concept of a flavor neutrinos propagator.

Our paper is organized in the following way. First, in Sec.~\ref{sec:QFTNUOSC},
we remind the basics of the QFT approach to neutrino flavor oscillations.
The analysis of the attempts to derive flavor neutrinos propagators in different variants
of the QFT is provided in Secs.~\ref{sec:OPER} and~\ref{sec:PATH}.
We construct the effective propagator of flavor neutrinos in Sec.~\ref{sec:EFFPROPVAC}.
The results of Sec.~\ref{sec:EFFPROPVAC} are applied to neutrino
flavor oscillations in vacuum in Sec.~\ref{sec:NUOSCVAC}. In Sec.~\ref{sec:PROPMATT},
we analyze the possibility to apply the effective propagators for
neutrino oscillations in background matter. Finally, we conclude in
Sec.~\ref{sec:CONCL}. Some properties of the Poincar\'e algebra are
given in Appendix~\ref{sec:POINCALG}. In Appendix~\ref{sec:USEFULINT},
we provide the values of the integrals for the calculation of the
matrix element.

\section{QFT approach to neutrino oscillations}\label{sec:QFTNUOSC}

In this section, we briefly remind how to apply the QFT to describe neutrino
flavor oscillations.

In experiments involving neutrinos, one does not observe these particles
directly. In a source and a detector, we deal with charged lepton
counterparts of neutrinos, whereas neutrinos behave like virtual particles.
Therefore, if the number of charged leptons in a detector is different
from that in a source, we can say that virtual neutrinos oscillate
while traveling from a source to a detector. This model of neutrino
oscillations was proposed first in Ref.~\cite{Kob82} and then independently
rediscovered in Refs.~\cite{GiuKimLee93,GriSto96}.

\begin{figure}
\centering
\includegraphics[scale=1]{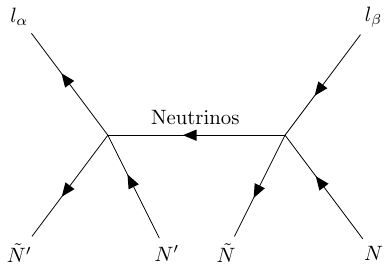}
\caption{The Feynman diagram corresponding to the $S$-matrix element in Eq.~(\ref{eq:Smatr}).}\label{fig:feyndiag}

\end{figure}

Within this model, one can rederive the QM probability of neutrino
oscillations which periodically varies with distance between a source
and a detector. Note that reproducing the QM result is practically
independent of the model of a source and a detector if reasonable
assumptions about the energies of incoming and outgoing particles
are made. Thus, following Ref.~\cite{Kob82}, we can approximate
a source and a detector by heavy nuclei which do not change their
positions in space. We depict the process of neutrino oscillations
schematically in Fig.~\ref{fig:feyndiag}.

We assume that neutrinos interact with charged leptons in frames of
the standard model, i.e. only left projections, e.g., $\nu_{\mathrm{L}}=\tfrac{1}{2}(1-\gamma^{5})\nu$,
of the operator valued wavefunctions are involved. Here, $\gamma^{5}=\mathrm{i}\gamma^{0}\gamma^{1}\gamma^{2}\gamma^{3}$
and $\gamma^{\mu}=(\gamma^{0},\bm{\gamma})$ are the Dirac matrices.
The analytical expression for the $S$-matrix element, corresponding
to Fig.~\ref{fig:feyndiag}, has the form,
\begin{equation}
S=-\frac{1}{2}\left(\sqrt{2}G_{\mathrm{int}}\right)^{2}\int\mathrm{d}^{4}x\mathrm{d}^{4}y\left\langle \tilde{N},\tilde{N}',l_{\alpha}\left|\mathrm{T}\left\{ j_{\mu}^{\dagger}(x)J^{\mu}(x)j^{\nu}(y)J_{\nu}^{\dagger}(y)\right\} \right|N,N',l_{\beta}\right\rangle ,\label{eq:Smatr}
\end{equation}
where
\begin{equation}
j_{\mu}=\sum_{\lambda}\bar{\nu}_{\lambda\mathrm{L}}\gamma_{\mu}l_{\lambda\mathrm{L}},\label{eq:lepcurr}
\end{equation}
is operator valued the leptonic current, $J^{\mu}$ is the nuclear
current operator, and $G_{\mathrm{int}}$ is the coupling constant,
and $\mathrm{T}$ is the symbol of the chronological ordering. We
assumed earlier that the nuclei $N$ and $\tilde{N}$, as well as
$N'$ and $\tilde{N}'$, are rather heavy. Thus, one gets that
\begin{equation}
\left\langle \tilde{N}'\left|J_{\mu}(x_{0},\mathbf{x})\right|N'\right\rangle \propto\delta_{\mu0}\delta(\mathbf{x}-\mathbf{x}_{2}),\quad\left\langle \tilde{N}\left|J_{\nu}^{\dagger}(y_{0},\mathbf{y})\right|N\right\rangle \propto\delta_{\nu0}\delta(\mathbf{y}-\mathbf{x}_{1}),\label{eq:nuclcurr}
\end{equation}
where $\mathbf{x}_{1}$ and $\mathbf{x}_{2}$ are the positions of
a source and a detector. Then, assuming that incoming and outgoing
charged leptons are plane waves, we obtain that
\begin{equation}
\left\langle l_{\alpha}\left|\mathrm{T}\left\{ j_{0}^{\dagger}(x)j_{0}(y)\right\} \right|l_{\beta}\right\rangle =e^{\mathrm{i}p_{\alpha}x-\mathrm{i}p_{\beta}y}\bar{u}_{\alpha}(p_{\alpha})\gamma_{0}^{\mathrm{L}}\left\langle 0\left|\mathrm{T}\left\{ \nu_{\alpha}(x)\bar{\nu}_{\beta}(y)\right\} \right|0\right\rangle \gamma_{0}^{\mathrm{L}}u_{\beta}(p_{\beta}),\label{eq:lepav}
\end{equation}
where $p_{\beta}^{\mu}=(E_{\beta},\mathbf{p}_{\beta})$ and $p_{\alpha}^{\mu}=(E_{\alpha},\mathbf{p}_{\alpha})$
are the four momenta of incoming and outgoing leptons, and $u_{\beta,\alpha}$
are their spinors.

There is a temptation to declare 
\begin{equation}
\mathrm{i}S_{\alpha\beta}(x-y)=\left\langle 0_{\mathrm{F}}\left|\mathrm{T}\left\{ \nu_{\alpha}(x)\bar{\nu}_{\beta}(y)\right\} \right|0_{\mathrm{F}}\right\rangle \label{eq:Salphabetadef}
\end{equation}
as the propagators of flavor neutrinos. However, we shall see shortly
in Sec.~\ref{sec:OPER} that the definition in Eq.~(\ref{eq:Salphabetadef})
contradicts the basic principles of QFT.

To avoid dealing with propagators of flavor neutrinos, one defines
the neutrino mass eigenstates $\psi_{a}$, which have definite masses
$m_{a}$, by
\begin{equation}
\nu_{\lambda}=\sum_{a}U_{\lambda a}\psi_{a},\label{eq:nuUpsi}
\end{equation}
where $(U_{\lambda a})$ is the mixing matrix. The propagators of
mass eigenstates in vacuum $S_{ab}=\delta_{ab}S_{a}$, where $S_{a}(x-y)=-\mathrm{i}\left\langle 0\left|\mathrm{T}\left\{ \psi_{a}(x)\bar{\psi}_{a}(y)\right\} \right|0\right\rangle $
is the propagator of a single massive neutrino (see also Eq.~(\ref{eq:Sadef})
below), which is well defined in the QFT.

Eventually, we cast Eq.~(\ref{eq:Smatr}) to
the form,
\begin{equation}
S=-2\pi G_{\mathrm{int}}^{2}\delta(E_{\alpha}-E_{\beta})e^{-i\mathbf{p}_{\alpha}\mathbf{x}_{2}+i\mathbf{p}_{\beta}\mathbf{x}_{1}}\mathrm{i}\mathcal{M}_{\beta\to\alpha},
\end{equation}
where
\begin{equation}
\mathcal{M}_{\beta\to\alpha}=\bar{u}_{\alpha}(p_{\alpha})\gamma_{0}^{\mathrm{L}}\sum_{a}U_{\alpha a}U_{\beta a}^{*}\left(\int\frac{\mathrm{d}^{3}q}{(2\pi)^{3}}e^{\mathrm{i}\mathbf{qL}}S_{a}(E,\mathbf{q})\right)\gamma_{0}^{\mathrm{L}}u_{\beta}(p_{\beta}),\label{eq:matrelgen}
\end{equation}
is the matrix element, $E=(E_{\alpha}+E_{\beta})/2$ is the mean energy
of incoming and outgoing leptons, $\mathbf{L}=\mathbf{x}_{2}-\mathbf{x}_{1}$
is the vector connecting a source and a detector, and $S_{a}(q_{0},\mathbf{q})$
is the 4D Fourier image of $S_{a}(x)$. Based on Eq.~(\ref{eq:matrelgen}),
one can reproduce the probability of neutrino flavor oscillations,
$P_{\nu_{\beta}\to\nu_{\alpha}}\propto|\mathcal{M}_{\beta\to\alpha}|^{2}$;
cf. Refs.~\cite{Kob82,GiuKimLee93,GriSto96}.

As one can see, the milestone of the QFT treatment of neutrino oscillations
is the propagators of virtual neutrinos. For massive neutrinos the
propagators are well defined. Despite we shall see in Secs.~\ref{sec:OPER}
and~\ref{sec:PATH} that the naive definition of a flavor neutrinos
propagator is invalid, still, one can construct the effective propagator
of flavor eigenstates in a consistent way. This kind of effective
propagator turns out to be useful for flavor oscillations.

\section{Propagators in the operator QFT}\label{sec:OPER}

In this section, we demonstrate how a propagator of a fermionic field
is constructed within the operator formalism in the QFT. We also demonstrate
that a properly defined propagator is valid only for neutrinos with
certain masses.

The QFT is based on the hypothesis of a weak coupling constant between
interacting fields. Thus, a process involving elementary particles
looks like a scattering. The probability of such a process is based
on the $S$-matrix operator,
\begin{equation}
S=\mathrm{T}\exp\left[-\mathrm{i}\int_{-\infty}^{+\infty}H_{\mathrm{I}}(t)\mathrm{d}t\right],\label{eq:Smatrdef}
\end{equation}
which connects the state vectors at $t=\mp\infty$, 
\begin{equation}
\left|\Phi_{+\infty}\right\rangle =S\left|\Phi_{-\infty}\right\rangle .\label{eq:PhiSPhi}
\end{equation}
In Eqs.~(\ref{eq:Smatrdef}) and~(\ref{eq:PhiSPhi}), the interaction
Hamiltonian $H_{\mathrm{I}}$ and the state vectors $\left|\Phi_{\pm\infty}\right\rangle $
are in the interaction representation. That is, the fields operators
$u(x)$, which $H_{\mathrm{I}}$ depends on, are considered to be
in the Heisenberg representation for noninteracting fields.

Now, we assume that $H_{\mathrm{I}}$ contains fermionic field $\psi_{a}$.
Expanding $S$ in Eq.~(\ref{eq:Smatrdef}) over a coupling constant
and using the Wick theorem, one gets the following correlator:
\begin{equation}
\mathrm{i}S_{a}(x-y)=\left\langle 0\left|\mathrm{T}[\psi_{a}(x)\bar{\psi}_{a}(y)]\right|0\right\rangle ,\label{eq:Sadef}
\end{equation}
if the field $\psi_{a}$ is not among incoming and outgoing particles,
i.e. if it is a virtual particle. The function $S_{a}(x)$ in Eq.~(\ref{eq:Sadef})
is called the causal Green function or the propagator.

We have already mentioned that $\psi_{a}$ is the noninteracting field
in the Heisenberg representation. If $\psi_{a}$ has the definite
mass $m_{a}$, it obeys the homogeneous Dirac equation,
\begin{equation}
(\mathrm{i}\gamma^{\mu}\partial_{\mu}-m_{a})\psi_{a}=0.\label{eq:Direqpsia}
\end{equation}
Based on the general solution of Eq.~(\ref{eq:Direqpsia}), we rewrite
Eq.~(\ref{eq:Sadef}) in the form,
\begin{equation}
S_{a}(x)=\int\frac{\mathrm{d}^{4}p}{(2\pi)^{4}}\frac{e^{-\mathrm{i}px}}{\not p-m_{a}+\mathrm{i}0},\label{eq:Saint}
\end{equation}
where $\mathrm{i}0$ is a small imaginary term. One can see that $S_{a}$
in Eq.~(\ref{eq:Saint}) obeys the equation,
\begin{equation}
(\mathrm{i}\gamma^{\mu}\partial_{\mu}-m_{a})S_{a}=\delta(x).\label{eq:DireqSa}
\end{equation}
However, we stress that, in the operator formulation of the QFT, the definition
of the propagator is in Eq.~(\ref{eq:Sadef}) rather than in Eq.~(\ref{eq:DireqSa}).
The term $\mathrm{i}0$ in Eq.~(\ref{eq:Saint}) brings sense to
the formal solution of Eq.~(\ref{eq:DireqSa}) since it instructs
how to bypass the poles in the Green function.

Now, let us consider the situation when we deal with mixed flavor
neutrinos having the Lagrangian,
\begin{equation}
\mathcal{L}=\sum_{\alpha\beta=e,\mu,\dotsc}\bar{\nu}_{\alpha}(\mathrm{i}\gamma^{\mu}\partial_{\mu}-m_{\alpha\beta})\nu_{\beta},\label{eq:Lagrfl}
\end{equation}
where the mass matrix $(m_{\alpha\beta})$ is nondiagonal. We make
an attempt to construct the propagator of $\nu_{\alpha}$.

There is a temptation to define the propagator of flavor neutrinos
as in Eq.~(\ref{eq:Salphabetadef}), analogously to Eq.~(\ref{eq:Sadef}).
Note that the definition in Eq.~(\ref{eq:Salphabetadef}) was used
in Ref.~\cite{CarChu99}. However, since the fields $\nu_{\alpha}$
do not obey free Dirac equations like in Eq.~(\ref{eq:DireqSa})
(flavor neutrinos interact between themselves even at $t\to\pm\infty$),
the jump from Eq.~(\ref{eq:Salphabetadef}) to
\begin{equation}
S_{\alpha\beta}(x)=\int\frac{\mathrm{d}^{4}p}{(2\pi)^{4}}\frac{e^{-\mathrm{i}px}}{\not p-(m_{\alpha\beta})+\mathrm{i}0},\label{eq:Salphabetaint}
\end{equation}
which was used for example in Refs.~\cite{KriSim23,KovSim24}, is
ambiguous.

One faces an additional hidden problem with Eq.~(\ref{eq:Salphabetadef}).
The whole Fock space $\left|\Phi_{\mathrm{F}}\right\rangle $, as
well as the vacuum state $\left|0_{\mathrm{F}}\right\rangle $ in
particular, are not defined for flavor neutrinos. In the axiomatic
formulation of the operator QFT, both coordinates $x^{\mu}$, and
operator valued wavefunctions $u(x)$, and state vectors $\left|\Phi\right\rangle $
are transformed under certain representations of the Poincar\'e group~\cite[p.~87]{BogShi80},
\begin{align}
x^{\mu} & \to x^{\prime\mu}=\hat{L}(\bm{\omega})x^{\mu}=L_{\nu}^{\mu}x^{\nu}+a^{\mu},\\
u(x) & \to u'(x')=\hat{\Lambda}(\bm{\omega})u(x),\label{eq:Poincu}\\
\left|\Phi\right\rangle  & \to\left|\Phi'\right\rangle =\hat{U}(\bm{\omega})\left|\Phi\right\rangle ,\label{eq:PoincPhi}
\end{align}
where $\bm{\omega}=(\omega_{\nu}^{\mu},a^{\mu})$ are the parameters
of the transformation.

The consistency of Eqs.~(\ref{eq:Poincu}) and~(\ref{eq:PoincPhi})
together with the unitarity of $\hat{U}$ result in the following
changes of $u$ and $\left|\Phi\right\rangle $ under the infinitesimal
transformation~\cite[p.~89]{BogShi80}:
\begin{align}
u'(x) & =\left(1+\mathrm{i}p_{\mu}a^{\mu}+\frac{\mathrm{i}}{2}m_{\mu\nu}\omega^{\mu\nu}\right)u(x),\nonumber \\
\left|\Phi'\right\rangle  & =\left(1+\mathrm{i}P_{\mu}a^{\mu}+\frac{\mathrm{i}}{2}M_{\mu\nu}\omega^{\mu\nu}\right)\left|\Phi\right\rangle ,\label{eq:Phiinftrans}
\end{align}
where $p_{\mu}=\mathrm{i}\partial_{\mu}$ and $m_{\mu\nu}=\mathrm{i}\left(x_{\mu}\partial_{\nu}-x_{\nu}\partial_{\mu}\right)$
are the generators of the boosts and the Lorentz rotations in the
coordinate representation. The form of the operators $P_{\mu}$ and
$M_{\mu\nu}$ depends on the specific implementation of the state
vector or, mathematically speaking, on the representation of the Poincar\'e
group in the state vector space. Some characteristics of the Poincar\'e
algebra are summarized in Appendix~\ref{sec:POINCALG}.

A particle is considered to be elementary if the representation of
the Poincar\'e group for $\left|\Phi\right\rangle $ is irreducible.
Any irreducible representation is characterized by the values of two
Casimir operators~\cite[pp.~60--63]{Ryd96},
\begin{equation}
P^{2}=m^{2},\quad W^{2}=-m^{2}s(s+1),\label{eq:Casimir2}
\end{equation}
which commute with all the generators $P_{\mu}$ and $M_{\mu\nu}$.
In Eq.~(\ref{eq:Casimir2}), $m$ is the particle mass, $s$ is the
particle spin, and $W^{\mu}$ is the Pauli--Lubanski vector given
in Eq.~(\ref{eq:PaulLub}).

One can see in Eqs.~(\ref{eq:Phiinftrans}) and~(\ref{eq:Casimir2})
that the Poincar\'e invariance of the whole quantum system is guaranteed
only for particles with certain masses. Of course, here, we assume
that $m>0$. That is, the construction of a state vector for flavor
neutrinos is ambiguous.

The vacuum state, which is, roughly speaking, the absence of everything,
can be defined only for massive particles in frames of the axiomatic
QFT. First, we note that the solution of a wave equation for a field
$u(x)$ has positive and negative frequency parts~\cite[pp.~92--94]{BogShi80},
\begin{align}
u_{\pm}(x) & =\int_{k_{0}>0}\mathrm{d}^{3}ke^{\pm\mathrm{i}kx}\delta(k^{2}-m^{2})\tilde{u}(\pm k)\label{eq:upm}
\end{align}
where $k_{0}=\sqrt{\mathbf{k}^{2}+m^{2}}$. Note that Eq.~(\ref{eq:upm})
is valid for any spin. The vacuum state $\left|0\right\rangle $ is
defined in such a way that $u_{-}(x)\left|0\right\rangle =0$ for
any types of particles. Alternatively, we can write it down in the
momentum representation $u_{-}(k)\left|0\right\rangle =0$. Here it
is important that $k^{2}=m^{2}$, i.e. the vacuum state is defined
for particles with certain masses. Therefore, the symbol $\left|0_{\mathrm{F}}\right\rangle $
used in Eq.~(\ref{eq:Salphabetadef}) is ill-defined.

\section{Propagators in path integral QFT}\label{sec:PATH}

The alternative approach to the formulation of QFT is based on the
path integral method. It allows one to derive all the quantities in
the explicit Lorentz invariant form. In this section, we make an attempt
to obtain the propagator for flavor neutrinos and demonstrate that
this derivation is invalid.

We rewrite the Lagrangian in Eq.~(\ref{eq:Lagrfl}) as
\begin{equation}
\mathcal{L}=\bar{N}(\mathrm{i}\gamma^{\mu}\partial_{\mu}-M)N,\label{eq:LagrflN}
\end{equation}
where $M=(m_{\alpha\beta})$ is the nondiagonal mass matrix and $N=(\nu_{e},\nu_{\mu},\dotsc)$
is the multiplet incorporating all neutrino types.

We can try to describe the dynamics of the system in Eq.~(\ref{eq:LagrflN})
using the generating functional (see, e.g., Ref.~\cite[p.~59]{FadSla80}),
\begin{equation}
Z[\bar{\xi},\xi]\propto\int\exp\left[\mathrm{i}\int\mathrm{d}^{4}x\left(\mathcal{L}+\bar{\xi}N+\bar{N}\xi\right)\right]\prod_{x}\mathrm{d}\bar{N}\mathrm{d}N,\label{eq:Zdef}
\end{equation}
in the presence of the external source $\xi$, which has the same
dimension as $N$. The path integral in Eq.~(\ref{eq:Zdef}) equals
(up to a constant) to the exponent in the integrand taken on the equations
of motion (see, e.g., Ref.~\cite[p.~54]{FadSla80}), i.e. $\left(\tfrac{\delta\mathcal{L}}{\delta\bar{N}}\right)_{\mathrm{L}}+\xi=0$
etc. Here, we take the left derivative with respect to the Grassmann
variable $\bar{N}$.

Eventually, one gets that $Z$ takes the form,
\begin{equation}
Z\propto\exp\left[-\mathrm{i}\int\mathrm{d}^{4}x\mathrm{d}^{4}y\bar{\xi}(x)(\mathrm{i}\gamma^{\mu}\partial_{\mu}-M)^{-1}\xi(y)\right].\label{eq:Zres}
\end{equation}
The Green function for flavor neutrinos is obtained by differentiating
Eq.~(\ref{eq:Zres}) with respect to $\xi$ and $\bar{\xi}$,
\begin{equation}
G(x-y)=-\left.\left(\frac{\delta}{\mathrm{i}\delta\xi(y)}\right)_{\mathrm{R}}\left(\frac{\delta}{\mathrm{i}\delta\bar{\xi}(x)}\right)_{\mathrm{L}}Z\right|_{\xi=\bar{\xi}=0}=-\frac{\mathrm{i}}{\mathrm{i}\gamma^{\mu}\partial_{\mu}-M}.\label{eq:Greenfunflav}
\end{equation}
Analogous method was used in Ref.~\cite[pp.~296--297]{FukYan03}
to get the propagators of Majorana neutrinos.

One can see that the Green function in Eq.~(\ref{eq:Greenfunflav})
formally coincides with a propagator of flavor neutrinos used in Refs.~\cite{KriSim23,KovSim24}
(see also Refs.~\cite{CarChu99,AkhWil13}, where the propagators
of flavor neutrinos in background matter were considered). However,
Eq.~(\ref{eq:Greenfunflav}) is meaningless since it is divergent.
To bring the sense to Eq.~(\ref{eq:Greenfunflav}) one should indicate
how to bypass the poles in this Green function.

We mention that the expression for $Z$ in Eq.~(\ref{eq:Zdef}) is
valid only if we request that $N$ obeys the boundary conditions at
$t\to\pm\infty$ (see, e.g., Ref.~\cite[p.~59]{FadSla80}),
\begin{equation}
N_{\text{out,in}}^{(\alpha)}(x)\propto\int\mathrm{d}^{3}k\left.N_{\mp}(k)e^{\pm\mathrm{i}kx}\right|_{k_{0}=\sqrt{\mathbf{k}^{2}+m_{\nu_{\alpha}}^{2}}}.\label{eq:boundcond}
\end{equation}
In this situation, the inverse operator $(\mathrm{i}\gamma^{\mu}\partial_{\mu}-M)^{-1}$
in Eq.~(\ref{eq:Zres}) acquires the small imaginary part $M\to M-\mathrm{i}0$
and the Green function in Eq.~(\ref{eq:Greenfunflav}) becomes the
propagator.

One can see in Eq.~(\ref{eq:boundcond}) that the boundary conditions
are given for particles with certain energies $\sqrt{\mathbf{k}^{2}+m_{\nu_{\alpha}}^{2}}$
and masses $m_{\nu_{\alpha}}$. Hence, the matrix $M$ should be diagonal.
It means that, in the path integral formalism, the Green function
in Eq.~(\ref{eq:Greenfunflav}), which is derived formally, is converted
to the propagator for the neutrino mass eigenstates rather than for
flavor neutrinos.

\section{Effective propagator of flavor neutrinos in vacuum}\label{sec:EFFPROPVAC}

We showed in Secs.~\ref{sec:OPER} and~\ref{sec:PATH} that defining
propagators of mixed flavor neutrinos in Eqs.~(\ref{eq:Salphabetadef})
and~(\ref{eq:Greenfunflav}) contradicts the basic principles of
QFT. Nevertheless, in this section, we demonstrate that one can still
construct an effective propagator of flavor neutrinos.

Let us consider two neutrino flavors, e.g., $\nu_{e}$ and $\nu_{\mu}$,
and rewrite the Lagrangian in Eq.~(\ref{eq:Lagrfl}) as $\mathcal{L}=\mathcal{L}_{0}-V$,
where
\begin{equation}
\mathcal{L}_{0}=\sum_{\alpha=e,\mu}\bar{\nu}_{\alpha}(\mathrm{i}\gamma^{\mu}\partial_{\mu}-m_{\alpha})\nu_{\alpha},\quad V=m_{\mu e}\bar{\nu}_{\mu}\nu_{e}+\text{h.c.}\label{eq:L0V}
\end{equation}
In Eq.~(\ref{eq:L0V}), we explicitly separated the terms diagonal
in flavors, $\mathcal{L}_{0}$, and the flavor mixing term $V$. We
shall consider $V$ as perturbation pretending that $m_{\mu e}\ll m_{e,\mu}$.
In practice, $V$ does not behave like an external field approaching
to zero at remote space-like hypersurfaces. Hence, to account for
the contribution of $V$ to the evolution of the system properly,
we have to sum all the terms in the perturbative series.

The formalism for finding the propagators in the system with an external
field, which mixes different particle types, was developed in Refs.~\cite{Dvo25,Dvo26a,Dvo26b}.
It is based on solving of the Dyson-like equations for dressed propagators.

\begin{figure}
\centering
\includegraphics[scale=1]{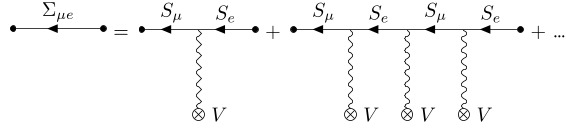}
\caption{The Feynman diagrams corresponding to the series in Eq.~(\ref{eq:Sigmaemuser})
which represent the dressed effective propagator $\Sigma_{\mu e}$.
The bare propagators $S_{e,\mu}$ are given in Eq.~(\ref{eq:Semu}).
The external potential is $V=m_{\mu e}$.}\label{fig:pertser}
\end{figure}

Let us consider, e.g., the propagator $\Sigma_{\mu e}$ which is responsible
for the transition $\nu_{e}\to\nu_{\mu}$. Based on Eq.~(\ref{eq:L0V}),
one gets that the perturbative series for $\Sigma_{\mu e}$ has the
form,
\begin{equation}
\Sigma_{\mu e}=S_{\mu}VS_{e}+S_{\mu}VS_{e}VS_{\mu}VS_{e}+\dotsb,\label{eq:Sigmaemuser}
\end{equation}
where
\begin{equation}
S_{e,\mu}=\frac{1}{\not p-m_{e,\mu}},\label{eq:Semu}
\end{equation}
are the Fourier images of the diagonal flavor propagators $\mathrm{i}S_{e,\mu}(x-y)=\left\langle 0_{\mathrm{e,\mu}}\left|\mathrm{T}[\nu_{e,\mu}(x)\bar{\nu}_{e,\mu}(y)]\right|0_{e,\mu}\right\rangle $.
Note that, now, the flavor vacua $\left|0_{e,\mu}\right\rangle $
are well defined since we construct the perturbation theory based
on $\mathcal{L}_{0}$ which is diagonal in flavors. In Eq.~(\ref{eq:Semu}),
we take that $m_{e,\mu}=m_{e,\mu}-\mathrm{i}0$ to bypass the poles
in the propagators. The Feynman diagrams representing the series in
Eq.~(\ref{eq:Sigmaemuser}) are shown in Fig.~\ref{fig:pertser}.

The result of the summation of the series in Eq.~(\ref{eq:Sigmaemuser})
can be written down as the Dyson-like equation,
\begin{equation}
(S_{\mu}VS_{e})^{-1}-V=\Sigma_{\mu e}^{-1}.\label{eq:Dyseq}
\end{equation}
Using Eq.~(\ref{eq:Semu}), one can solve Eq.~(\ref{eq:Dyseq})
exactly,
\begin{equation}
\Sigma_{\mu e}=\frac{m_{\mu e}(m_{e}+m_{\mu})\left(\not p+\frac{p^{2}+m_{e}m_{\mu}-m_{\mu e}^{2}}{m_{e}+m_{\mu}}\right)}{p^{4}-p^{2}(m_{e}^{2}+m_{\mu}^{2}+2m_{\mu e}^{2})+(m_{e}m_{\mu}-m_{\mu e}^{2})^{2}}.\label{eq:Sigmamue}
\end{equation}
Analogously, one can obtain that $\Sigma_{e\mu}=\Sigma_{\mu e}$.

\section{Neutrino flavor oscillations in vacuum}\label{sec:NUOSCVAC}

In this section, we apply the effective propagator, found in Sec.~\ref{sec:EFFPROPVAC},
for neutrino flavor oscillations in vacuum.

Based on Eq.~(\ref{eq:Sigmamue}), we rewrite the matrix element
for neutrino oscillations in Eq.~(\ref{eq:matrelgen}) in the form,
\begin{equation}
\mathcal{M}_{e\to\mu}=\bar{u}_{\mu}(p_{\mu})\gamma_{0}^{\mathrm{L}}\left(\int\frac{\mathrm{d}^{3}q}{(2\pi)^{3}}e^{\mathrm{i}\mathbf{qL}}\Sigma_{\mu e}(E,\mathbf{q})\right)\gamma_{0}^{\mathrm{L}}u_{e}(p_{e}).\label{eq:Memu}
\end{equation}
Then, we assume that $\mathbf{L}=L\mathbf{e}_{z}$ and use the approximation
of ultrarelativistic charged leptons propagating along the $z$-axis.
In this situation, $u_{e}=u_{\mu}=(0,0,0,1)^{\mathrm{T}}$. Then,
we use the identity $\bar{u}_{\mu}\gamma_{0}^{\mathrm{L}}(\not q+\tilde{m})\gamma_{0}^{\mathrm{L}}u_{e}=(E+q_{z})$
which is valid for any scalar $\tilde{m}$. When we parameterize the
neutrino three momentum as $\mathbf{q}=\rho\mathbf{e}_{\rho}+z\mathbf{e}_{z}$,
the matrix element in Eq.~(\ref{eq:Memu}) takes the form,
\begin{align}
\mathcal{M}_{e\to\mu}= & \frac{m_{\mu e}(m_{e}+m_{\mu})}{(2\pi)^{2}}\int_{0}^{\infty}\rho\mathrm{d}\rho\int_{-\infty}^{+\infty}\mathrm{d}ze^{\mathrm{i}zL}\nonumber \\
 & \times\frac{E+z}{(E^{2}-z^{2}-\rho^{2})^{2}-(E^{2}-z^{2}-\rho^{2})(m_{e}^{2}+m_{\mu}^{2}+2m_{\mu e}^{2})+(m_{e}m_{\mu}-m_{\mu e}^{2})^{2}}.\label{eq:Memudrdz}
\end{align}
The poles of the integrand in Eq.~(\ref{eq:Memudrdz}) are 
\begin{equation}
z_{1,2}^{2}=\rho_{1,2}^{2}-\rho^{2}+\mathrm{i}0,\label{eq:z12}
\end{equation}
where $\rho_{1,2}^{2}=E^{2}-m_{1,2}^{2}$ and 
\begin{equation}
m_{1,2}=\frac{1}{2}\left(m_{e}+m_{\mu}\mp\sqrt{(m_{e}-m_{\mu})^{2}+4m_{\mu e}^{2}}\right).\label{eq:m12}
\end{equation}
In Eq.~(\ref{eq:z12}), we take into account our convention about
the imaginary parts of $m_{e,\mu}$ in Eq.~(\ref{eq:Semu}). If we
assume that $m_{2}>m_{1}$, then $\rho_{2}<\rho_{1}$.

The result of the integration over $z$ and $\rho$ in Eq.~(\ref{eq:Memudrdz})
reads
\begin{equation}
\mathcal{M}_{e\to\mu}=\frac{m_{\mu e}(m_{e}+m_{\mu})E}{4\pi(m_{2}^{2}-m_{1}^{2})L}\left[e^{\mathrm{i}\rho_{1}L}\left(1+\frac{\rho_{1}}{E}+\frac{\mathrm{i}}{EL}\right)-e^{\mathrm{i}\rho_{2}L}\left(1+\frac{\rho_{2}}{E}+\frac{\mathrm{i}}{EL}\right)\right].\label{eq:Memures}
\end{equation}
When we integrate over $z$, we take into account that $L>0$. Hence,
we close the contour in the upper half-plane. Using Eq.~(\ref{eq:z12}),
one gets that $z_{1,2}=\sqrt{\rho_{1,2}^{2}-\rho^{2}}+\mathrm{i}0$
if $\rho<\rho_{1,2}$, and $z_{1,2}=\mathrm{i}\sqrt{\rho^{2}-\rho_{1,2}^{2}}$
if $\rho>\rho_{1,2}$. When we integrate over $\rho$, we have to
account for the all possible positions of the integration variable
with respect to $\rho_{1}$ and $\rho_{2}$. Moreover, we use the
values of the integrals in Eq.~(\ref{eq:usefint}).

We identify the masses $m_{1,2}$ in Eq.~(\ref{eq:m12}) with the
masses of mass eigenstates $\psi_{1,2}$. Thus, one should choose the
mixing matrix $(U_{\lambda a})$ in Eq.~(\ref{eq:nuUpsi}) to bring
the mass matrix $M_{\text{flavor}}$ to the diagonal form $M_{\text{mass}}$,
\begin{equation}
M_{\text{mass}}=U^{\dagger}M_{\text{flavor}}U,\quad M_{\text{mass}}=\text{diag}(m_{1},m_{2}),\quad M_{\text{flavor}}=\left(\begin{array}{cc}
m_{e} & m_{e \mu}\\
m_{\mu e} & m_{\mu}
\end{array}\right),
\end{equation}
where $m_{e \mu}=m_{\mu e}$. The mixing matrix can be represented
in the form,
\begin{equation}
U=\left(\begin{array}{cc}
\cos\theta & \sin\theta\\
-\sin\theta & \cos\theta
\end{array}\right).
\end{equation}
Here, $\theta$ is the vacuum mixing angle, which obeys the relation
\begin{equation}
\sin^{2}2\theta=\frac{4m_{\mu e}^{2}(m_{\mu}+m_{e})^{2}}{(\Delta m^{2})^{2}},\label{eq:thetarel}
\end{equation}
where $\Delta m^{2}=m_{2}^{2}-m_{1}^{2}$ is the mass squared difference.

As a rule, neutrino oscillations are studied at the macroscopic distance
$L\gg E^{-1}$. Thus, we can neglect the terms $\sim\mathcal{O}\left(\tfrac{1}{EL}\right)$
in Eq.~(\ref{eq:Memures}). Moreover, we consider ultrarelativistic
case, in which $E\gg m_{1,2}$. In this situation, $\rho_{1,2}\approx E-\frac{m_{1,2}^{2}}{2E}$
in Eq.~(\ref{eq:Memures}). The probability of oscillations $\nu_{e}\to\nu_{\mu}$
is $P_{\nu_{e}\to\nu_{\mu}}\propto|\mathcal{M}_{e\to\mu}|^{2}$. Using
Eqs.~(\ref{eq:Memures}) and~(\ref{eq:thetarel}), as well as the
above mentioned approximations, one gets that
\begin{equation}
P_{\nu_{e}\to\nu_{\mu}}\approx\sin^{2}2\theta\sin^{2}\left(\frac{\Delta m^{2}}{4E}L\right).\label{eq:Pemuvac}
\end{equation}
One can see that Eq.~(\ref{eq:Pemuvac}) reproduces the known QM
result for the transition probability of neutrino flavor oscillations
in vacuum.

\section{QFT for neutrino oscillations in matter}\label{sec:PROPMATT}

It was claimed in Ref.~\cite{KovSim24} that one can use the propagator
of flavor neutrinos in background matter,
\begin{equation}
S_{\alpha\beta}(x)=\int\frac{\mathrm{d}^{4}p}{(2\pi)^{4}}\frac{e^{-\mathrm{i}px}}{\not p-(m_{\alpha\beta})-\not V+\mathrm{i}0},\label{eq:Salphabetamatt}
\end{equation}
to describe neutrino oscillations in external background in frames
of QFT. In Eq.~(\ref{eq:Salphabetamatt}), $V^{\mu}=\text{diag}(V_{\nu_{e}}^{\mu},V_{\nu_{\mu}}^{\mu},\dots)$
is the multiplet of the effective potentials for the neutrino interaction
with matter. It is important that $V^{\mu}$ is diagonal in neutrino
flavors, which is the feature of the standard model.

We showed in Secs.~\ref{sec:OPER} and~\ref{sec:PATH} that the propagator
of flavor neutrinos like that in Eq.~(\ref{eq:Salphabetamatt}) is
ill-defined. Now, we demonstrate that the effective propagator of flavor neutrinos in matter, which is analogous to that in Sec.~\ref{sec:EFFPROPVAC}, cannot be derived either.

Suppose
that we deal with Dirac neutrinos. We obtained the exact propagator
for an ultrarelativistic massive neutrino in matter in Ref.~\cite{Dvo26a},
\begin{equation}
S_{e,\mu}(p_{0},\mathbf{p})=-\frac{1}{(p_{0}-E_{e,\mu}^{(-)}-V_{e,\mu}/2+\mathrm{i}0)}\frac{(1-\gamma^{5})}{2}\left(\begin{array}{cc}
0 & \tfrac{1}{2}\left[1-\left(\bm{\sigma}\hat{\mathbf{p}}\right)\right]\\
\tfrac{1}{2}\left[1-\left(\bm{\sigma}\hat{\mathbf{p}}\right)\right] & 0
\end{array}\right),\label{eq:propdiagur}
\end{equation}
where $V_{e,\mu}=V_{e,\mu}^{0}\propto G_{\mathrm{F}}n_{\mathrm{eff}}$
are the effective potentials for electron and muon neutrinos interaction
with a nonmoving matter, $E_{e,\mu}^{(-)}=\sqrt{\left(|\mathbf{p}|+\tfrac{V_{e,\mu}}{2}\right)^{2}+m_{e,\mu}^{2}}$,
$\hat{\mathbf{p}}=\mathbf{p}/p$ is the unit vector along $\mathbf{p}$,
$\bm{\sigma}$ are the Pauli matrices, $G_{\mathrm{F}}$ is the Fermi
constant, and $n_{\mathrm{eff}}$ is the effective number density
of background fermions.

Using Eq.(\ref{eq:Sigmaemuser}), one can see that all terms in the
perturbative series vanish if we take $S_{e,\mu}$ in Eq.~(\ref{eq:propdiagur})
and $V=m_{\mu e}$. Indeed, the propagator in Eq.~(\ref{eq:propdiagur})
is based on the exact solution of a Dirac equation for a left-handed
neutrino, whereas the nondiagonal mass term $\propto m_{\mu e}\bar{\nu}_{\mu}\nu_{e}+\text{h.c.}=m_{\mu e}\bar{\nu}_{\mu\mathrm{L}}\nu_{e\mathrm{R}}+\dotsb$
mixes different particle chiralities. That is why, $\Sigma_{\mu e}=S_{\mu}m_{\mu e}S_{e}+\dotsb=0$.
One can obtain the same result by solving the Dyson Eq.~(\ref{eq:Dyseq}).
In this situation, one should regularize Eq.~(\ref{eq:propdiagur})
analogously to Ref.~\cite{Dvo26a}.

Therefore, we cannot define an effective propagator for flavor neutrinos
interacting with background matter in a QFT consistent way. The only
way to study neutrino flavor oscillations in matter in frames of the QFT
is to use propagators of mass eigenstates as in Ref.~\cite{Dvo26a}
for Dirac neutrinos or analogously to Ref.~\cite{Dvo25} for Majorana
ones.

\section{Conclusion}\label{sec:CONCL}

In this work, we have analyzed the application of QFT for the description
of neutrino flavor oscillations. In this formalism, neutrinos are
taken to be virtual particles. Thus, in the calculation of the matrix
element, one should have the propagators of these neutrinos. Nevertheless,
in Refs.~\cite{CarChu99,AkhWil13,KriSim23,KovSim24}, where neutrino
flavor oscillations were studied in vacuum and in background matter,
the propagators of flavor eigenstates were considered.

We have demonstrated that the construction of a propagator of a field
with an indefinite mass, which a flavor neutrino is, violates the
basic principles of QFT. It was shown in Sec.~\ref{sec:OPER} for
the operator formulation of QFT. In this situation, both operator
valued wavefunctions and vectors of states should change under a Poincar\'e
group transformation. However, if a flavor neutrino has neither a
definite mass nor a definite energy, no element of a Fock space, which
transforms under an irreducible representation of the Poincar\'e group,
can be constructed. It should be noted that the problem with the Poincar\'e
invariance of the Fock space of flavor neutrinos was admitted in Ref.~\cite{BlaSma25}.
An attempt to solve this problem was made in Ref.~\cite{Lob19}. 

The alternative formulation of QFT is based on the path integral method.
It allows one to get the $S$-matrix in the explicitly Lorentz invariant
form. Nevertheless, we have demonstrated in Sec.~\ref{sec:PATH}
that the naive construction of a propagator of flavor neutrinos violates
the boundary conditions for these fields at $t\to\pm\infty$. These
conditions are inherent to bring sense to a Green function. In particular,
they point out how to bypass the poles in a Green function to convert
it to a propagator.

If we are in frames of the relativistic quantum mechanics, both mass
and flavor bases are equivalent, i.e. they are connected by a unitary
transformation. However, as we have shown in Secs.~\ref{sec:OPER}
and~\ref{sec:PATH}, using QFT imposes stronger constrains on a system,
which make it impossible to derive a propagator of flavor neutrino
is the same way as for mass eigenstates.

Despite the naive construction of a propagator of flavor neutrinos
is impossible, we have shown in Sec.~\ref{sec:EFFPROPVAC} that one
can build the effective propagator of these fields. For this purpose,
we have considered two flavor neutrinos, $\nu_{e}$ and $\nu_{\mu}$,
and have treated the nondiagonal element in the mass matrix as the
perturbation. Applying the formalism in Refs.~\cite{Dvo25,Dvo26a,Dvo26b},
we have shown that the dressed propagator $\Sigma_{\mu e}$, which
is responsible for $\nu_{e}\to\nu_{\mu}$ oscillations, obeys the
Dyson-like equation. We have solved this equation exactly. Note that
the analogous idea for neutrino flavor oscillations was considered
in Refs.~\cite{BlaSma25,Tur23}.

Then, in Sec.~\ref{sec:NUOSCVAC}, we have applied the results of
Sec.~\ref{sec:EFFPROPVAC} to describe neutrino flavor oscillations
in vacuum. Making reasonable assumptions on the properties of incoming
and outgoing particles, we have rederived the QM transition probability
for flavor oscillations.

Finally, in Sec.~\ref{sec:PROPMATT}, we have analyzed the possibility
to apply the developed formalism to neutrino oscillations in background
matter. We have demonstrated that an effective propagator of a flavor
neutrino in background matter is unlikely to be constructed provided
that we do not violate the basic principles of the QFT.

\section*{Acknowledgments}

I am thankful to Oleg Teryaev for fruitful discussions.

\appendix

\section{Properties of the Poincar\'e algebra}\label{sec:POINCALG}

The Poincar\'e algebra has ten generators $P_{\mu}$ and $M_{\mu\nu}=-M_{\nu\mu}$,
which obey the following commutation rules:
\begin{align}
[P_{\mu},P_{\nu}] & =0,\quad[P_{\mu},M_{\rho\sigma}]=\mathrm{i}(g_{\mu\rho}P_{\sigma}-g_{\mu\sigma}P_{\rho}),\nonumber \\{}
[M_{\mu\nu},M_{\rho\sigma}] & =\mathrm{i}(g_{\nu\rho}M_{\mu\sigma}-g_{\mu\rho}M_{\nu\sigma}+g_{\mu\sigma}M_{\nu\rho}-g_{\nu\sigma}M_{\mu\rho}),\label{eq:Poincgen}
\end{align}
Then, we define the Pauli--Lubanski vector
\begin{equation}
W_{\rho}=-\frac{1}{2}\varepsilon_{\rho\alpha\beta\lambda}M^{\alpha\beta}P^{\lambda}.\label{eq:PaulLub}
\end{equation}
Using Eq.~(\ref{eq:Poincgen}), one can show that $C_{1}=P_{\mu}P^{\mu}$
and $C_{2}=W_{\mu}W^{\mu}$, which are called the Casimir operators,
commute with all $P_{\mu}$ and $M_{\mu\nu}$.

The values of $C_{1,2}$ are $C_{1}=m^{2}$ and $C_{2}=-m^{2}s(s+1)$,
where $m>0$ is the particle mass and $s$ is the particle spin. If
$m=0$, $W_{\mu}=\lambda P_{\mu}$, where $\lambda$ is the particle
helicity. 

\section{Some useful integrals}\label{sec:USEFULINT}

In this Appendix, we provide some integrals used in Sec.~\ref{sec:NUOSCVAC}
to compute the matrix element. They are
\begin{align}
\int_{0}^{\rho_{0}}\rho\mathrm{d}\rho\frac{e^{\mathrm{i}L\sqrt{\rho_{0}^{2}-\rho^{2}}}}{\sqrt{\rho_{0}^{2}-\rho^{2}}} & =-\frac{\mathrm{i}}{L}(e^{\mathrm{i}\rho_{0}L}-1)\nonumber \\
\int_{0}^{\rho_{0}}\rho\mathrm{d}\rho e^{\mathrm{i}L\sqrt{\rho_{0}^{2}-\rho^{2}}} & =-\frac{\mathrm{i}}{L^{2}}e^{\mathrm{i}\rho_{0}L}(L\rho_{0}+\mathrm{i})-\frac{1}{L^{2}},\nonumber \\
\int_{\rho_{0}}^{\infty}\rho\mathrm{d}\rho\frac{e^{-L\sqrt{\rho^{2}-\rho_{0}^{2}}}}{\sqrt{\rho^{2}-\rho_{0}^{2}}} & =\frac{1}{L},\nonumber \\
\int_{\rho_{0}}^{\infty}\rho\mathrm{d}\rho e^{-L\sqrt{\rho^{2}-\rho_{0}^{2}}} & =\frac{1}{L^{2}}.\label{eq:usefint}
\end{align}
One can check the validity of Eq.~(\ref{eq:usefint}) by means of
the direct calculations.

\end{document}